\documentclass{doublecol-new}
\usepackage{graphicx}
\usepackage{natbib,stfloats}
\usepackage{colortbl}
\usepackage{rotating}

\theoremstyle{TH}{

}

\theoremstyle{THrm}{

}

\theoremstyle{THhit}{

}

\makeatletter
\def\theequation{\arabic{equation}}
\makeatother

\begin{document}%

\LRH{Nilo Serpa}

\RRH{Theoretical Count of Function Points for Non-Measurable Items}

\VOL{1}

\ISSUE{3/4}

\PUBYEAR{2010}

\BottomCatch

\title{Theoretical Count of Function Points for Non-Measurable Items}

\authorA{Nilo Serpa}

\affA{Instituto de Ci\^{e}ncias Exatas e Tecnologia,\\
 UNIP - Universidade Paulista,\\
 SGAS Quadra 913, s/n$^o$ - Conjunto B - Asa Sul - Bras\'{i}lia - DF, Brasil\\
 CEP 70390-130\\ e-mail: nilo@techsolarium.com\\
 POLITEC Global IT Services,\\
 SIG Quadra 4 - lote 173, Setor Gr\'{a}fico - Bras\'{i}lia - DF, Brasil\\
 CEP 70610-440}

\begin{abstract}
This paper studies and proposes an extended technique of function point counting for items classified as non-measurable. The main objective is to expand the conventional technique of counting to ensure that it comprises consistently the tasks involved in building portals and sites in general. In addition, it also applies to measure the cost of continued activities related to these web applications. The extended technique is potentially useful to measure several products associated with information systems, including periodicals publishable in intranets.
\end{abstract}

\KEYWORD{function points; intellectual effort; inertia of development; econophysics; Lagrangian dynamics; metrics}

\begin{bio}
Nilo Serpa is {\it Magister in Scientia}, in Astronomy, from the {\it Universidade do Brasil}, and Master of Business Administration from the {\it Funda\c{c}\~{a}o Getulio Vargas}, Brazil. His {\it Magister} Thesis was about applications of the inhomogeneous Lema\^{i}tre-Tolman cosmology with an original approach of weak gravitational lensing in that cosmology. He specializes in Management Process and Information Technology from the {\it Funda\c{c}\~{a}o Oswaldo Cruz}, Brazil. His training in Physics at the {\it Centro Brasileiro de Pesquisas F\'{i}sicas - CBPF -} ranges from Quantum Mechanics to Supergravity, including Gauge Field Theories, Topological Effects and Supersymmetry. He received his degree of Architect in the year 1981, now having thirty years experience in IT as a Project Manager and Development Manager, and fifteen years experience in Physics as researcher and professor. In the early Eighties, it was pupil of the semiologist Umberto Eco, experience that contributed for his interest on the semantics of the mathematical formalizations. He is Associate Professor of Physics, Software Engineering and Professional Ethics at the {\it Universidade Paulista}, Brazil, and Senior Development Manager at {\it POLITEC Global IT Services}, Brazil. He is Analyst of Training with great experience in e-learning, having worked in collaboration with some of the greatest names in the area of qualification of human resources, such as the Canadian Software Engineer John Franklin Arce. He is also Special Project Manager at the General Coordination of Information and Informatics of the Ministry of Work and Employment, Brazil. Having received a Senior training in Function Point Analysis, he has created a methodology named "Priority Point Analysis" to measure an IT Coordination by the problems it faces. He is author of the book {\it Revers\~{o}es Geopol\'{i}ticas: Geografia, F\'{i}sica e Filosofia na Sociedade Globalizada} (2002). His main works in Physics are "Thermodynamics of Diabetes Mellitus: the Physical Reasons of Obesity and Sedentarismus as Decisive Variables in Predictive Models", "New Lectures on Supergravity", "{\it El-Ni\~{n}o: Influ\^{e}ncia Exterior, Mat\'{e}ria Negra e Caudas Gravitacionais}", "The Counting of Galaxies from Type Ia Supernovae Rate" and "Modelling the Dynamics of the Work-Employment System with Predator-Prey Interactions". He is invited reviewer of the journal {\it Economic Modelling}. His areas of interest include Cosmology, Field Theory, Geopolitics, Econophysics and Information Technology.\break
\end{bio}

\maketitle\vfill\pagebreak

\maketitle
\thispagestyle{empty}


\maketitle

\newpage

\vspace*{-11pt}
{\it You can not imagine how everything is vague until try to do it accurately.}

\begin{flushright}
{\it Bertrand Russell}
\end{flushright}

\section{Introduction}	

The function point analysis (FPA) is a standardized technique for measuring
software development, aiming to establish a gauge of the software size based on the functionalities to be implemented, considering the viewpoint of the user \citep{16}. Some usage thoughts have been made to the extent that software systems are becoming more complex \citep{21}, such as:
\begin{enumerate}
\item "Function points are not a very good measure when sizing maintenance efforts
(fixing problems) or when trying to understand the performance issues. Much of
the effort associated with fixing problems (production fixes) is due to trying to
resolve and understand the problem (detective work)".
\item "FP analysis (FPA) is not useful to size Web Design. FPA is useful to size web
development, but not web design.[...] FPA is not useful in estimating the time necessary
to create graphics, images, page layouts, so on and so forth".
\end{enumerate}

During the eighties and until the beginning of the 21st century, a number of authors have discussed the metric procedures in vogue \citep{4}, \citep{12}, analyzing its applicability to object-oriented software \citep{25}, its advantages and disadvantages \citep{18}. Also, Kemerer (1993) studied the reliability of the FPA technique. Important contributions to understanding the complexity inherent to software engineering were brought by Indian school \citep{17},\citep{23}. More enthusiastic works about FPA appeared since 2000 \citep{13}. 

The function points measurement technique has generated much controversy since its dissemination as an ISO recognized tool to size information systems, both as regards its general purposes (does it measure productivity, size, complexity or functionality?) and in relation to mathematical rigor under the concept of metric. In particular, with respect to the latter, being the number of function points a dimensionless quantity, some authors claim that there was no way to analyze and seek information from numbers not associated with a reference system \citep{2}. This is not entirely true. In science there are many dimensionless numbers widely applied in several fields as hydrodynamics, geophysics, optics and others. A dimensionless metric, being independent of the reality beneath evaluation, is useful to compare two or more objects abstracting a lot of details of these objects, placing them in the same plane of observation and providing a perspective that would be inconceivable without a standardized approach. Perhaps the difficulty is to precise the mathematical structure and the semantics of the metric, that is, what the metric formally measures.

One of the major contractual problems faced both in the governmental sphere and in the context of private enterprise is the remuneration of the activities not measurable by the technique of function point analysis. Some devices have been adopted, but with high degree of arbitrariness, making the calculation uncertain and often unfair, and vulnerable to critical assessments of the organs of control. In addition, an arbitrary control of the estimates may be evidence of unprofessional management. Also, something more is missed out as observed by Lokan (2008):\\

{\it Function points are oriented towards data-strong systems, typified by business software. Processing in these systems is simple. Most effort goes into defining data structures. Not all systems fit this pattern. Scientific and engineering software is often function-strong: dominated by the internal processing required to transform inputs to outputs}.\\

The present model, although it stemmed from the need to measure the development effort of sites and portals,
aims to incorporate not just the typical non-measurable items but all function-strong software processes, including operational system migrations.

It is susceptible to broad questioning the measurement of all the tasks required in the design and development of web applications (especially portals), now applied to the Ministry of Work and Employment in Brazil (hereafter MTE), not only by the aesthetic and functional aspects but also by the technology of the software resources in use. It should also consider that there is here, as in other IT activities, a significant amount of intellectual effort that, while difficult to measure in any area, must be properly computed and paid even so by an indirect and approximate manner.

There are creative works published on the techniques of scoring for web systems \citep{1}, \citep{11}, and for situations in which {\it a priori} non-measurability is compensated by a technique to perform FPA based on the source code \citep{22}. Also, recent analytical studies call attention to important issues that remain unresolved and that should attract greater interest from the international community of IT \citep{15}. It draws attention, however, the fact that none of them is the proposition of a complete formalism that includes the representation of the intellectual effort, the hours worked and productivity of the tool applied, the most relevant of the few benchmarks for web systems. Indeed, the size in function points is not intended to measure productivity and development effort, but to measure software in terms of its functionality. Nevertheless, from the moment that we focus on the trinomial quality-cost-time it is imperative to compute the assets embedded in the engineering itself. It is noteworthy, as well pointed Aramo-Immonen {\it et al} (2011), the influence that cultural differences have on the productivity in globalized software engineering, which makes even more complex the challenge of managing productivity and costs, mainly in large corporations. It is the reality of this fact, coupled with the need to count function points in the management of IT services, which forces us to broaden the horizons of the technology, so that we can standardize our practices, regardless of culture or activity, preventing the proliferation of competitor procedures, and, consequently, the misconceptions which they can produce.

The name "function point" not even seem to fit the web development by the absence there of the classical counting elements. However, a web layout contains intrinsic functionalities beside menus, hyperlinks, validity checks of fill, etc.; the communicational function of the web layouts overlaps unequivocally to other aspects; whence the difficulty to score so abstract assignment; whence the request for the application of more formal techniques.

\begin{table*}[!p]
\begin{center}
\small
\caption{Tasks, complexities and durations}
\begin{tabular}{crrrrrrrrrrrr}

               \multicolumn{ 5}{c}{{\bf Task}} & {\bf Complexity} &     {\bf } & {\bf Optimist} & {\bf Pessimist} & {\bf Most likely} &      \\

    {\bf } &     {\bf } &     {\bf } &     {\bf } &     {\bf } &     {\bf } &     {\bf } &     {\bf } &     {\bf } &     {\bf } &     {\bf } &     \\

       \multicolumn{ 5}{l}{{\bf HTML conversion}} & {\bf Low} &            &    0:30:00 &    3:00:00 &    1:45:00 &                        \\

                                   \multicolumn{ 5}{l}{{\bf }} & {\bf Average} &            &    1:00:00 &    4:00:00 &    2:30:00 &                   \\

                                   \multicolumn{ 5}{l}{{\bf }} & {\bf High} &            &    1:30:00 &    6:00:00 &    3:45:00 &                       \\

    {\bf } &     {\bf } &     {\bf } &     {\bf } &     {\bf } &     {\bf } &            &            &            &            &                   \\

       \multicolumn{ 5}{l}{{\bf PDF conversion}} & {\bf Standard} &            &    0:30:00 &    0:30:00 &    0:30:00 &                \\

    {\bf } &     {\bf } &     {\bf } &     {\bf } &     {\bf } &     {\bf } &            &            &            &            &                    \\

    \multicolumn{ 5}{l}{{\bf Multimedia inserting}} & {\bf Low} &            &    0:30:00 &    2:00:00 &    1:15:00 &                      \\

                                   \multicolumn{ 5}{l}{{\bf }} & {\bf Average} &            &    1:00:00 &    3:30:00 &    2:15:00 &                   \\

                                   \multicolumn{ 5}{l}{{\bf }} & {\bf High} &            &    1:30:00 &    6:00:00 &    3:45:00 &                \\

    {\bf } &     {\bf } &     {\bf } &     {\bf } &     {\bf } &     {\bf } &            &            &            &            &                     \\

      \multicolumn{ 5}{l}{{\bf Image creation and treatment}} & {\bf Low} &            &    0:30:00 &    2:30:00 &    1:30:00 &                    \\

                                   \multicolumn{ 5}{l}{{\bf }} & {\bf Average} &            &    2:00:00 &    5:30:00 &    3:45:00 &                \\

                                   \multicolumn{ 5}{l}{{\bf }} & {\bf High} &            &    4:00:00 &   12:00:00 &    9:00:00 &                    \\

    {\bf } &     {\bf } &     {\bf } &     {\bf } &     {\bf } &     {\bf } &            &            &            &            &                       \\

              \multicolumn{ 5}{l}{{\bf Form creation}} & {\bf Low} &            &    0:30:00 &    4:00:00 &    2:30:00 &                   \\

                                   \multicolumn{ 5}{l}{{\bf }} & {\bf Average} &            &    2:00:00 &    8:00:00 &    6:00:00 &                 \\

                                   \multicolumn{ 5}{l}{{\bf }} & {\bf High} &            &    4:00:00 &   12:00:00 &    9:00:00 &                   \\

    {\bf } &     {\bf } &     {\bf } &     {\bf } &     {\bf } &     {\bf } &            &            &            &            &                     \\

  \multicolumn{ 5}{l}{{\bf Layout creation and development}} & {\bf Low} &            &    8:00:00 &    8:00:00 &    8:00:00 &                   \\

                                   \multicolumn{ 5}{l}{{\bf }} & {\bf Average} &            &   19:00:00 &   24:00:00 &   21:30:00 &                       \\

                                   \multicolumn{ 5}{l}{{\bf }} & {\bf High} &            &   38:00:00 &   40:00:00 &   39:00:00 &                      \\

    {\bf } &     {\bf } &     {\bf } &     {\bf } &     {\bf } &     {\bf } &            &            &            &            &                      \\

                \multicolumn{ 5}{l}{{\bf Layout adequation }} & {\bf Standard} &            &    3:00:00 &    8:00:00 &    5:30:00 &                  \\

    {\bf } &     {\bf } &     {\bf } &     {\bf } &     {\bf } &     {\bf } &            &            &            &            &                     \\

                 \multicolumn{ 5}{l}{{\bf Layout montage}} & {\bf Low} &            &    8:00:00 &    9:00:00 &    8:30:00 &                  \\

                                   \multicolumn{ 5}{l}{{\bf }} & {\bf Average} &            &   22:00:00 &   24:00:00 &   23:00:00 &                      \\

                                   \multicolumn{ 5}{l}{{\bf }} & {\bf High} &            &   36:00:00 &   40:00:00 &   38:00:00 &                    \\

    {\bf } &     {\bf } &     {\bf } &     {\bf } &     {\bf } &     {\bf } &            &            &            &            &                     \\

                 \multicolumn{ 5}{l}{{\bf Creation of Tables}} & {\bf Low} &            &    0:30:00 &    2:00:00 &    1:40:00 &                     \\

                                   \multicolumn{ 5}{l}{{\bf }} & {\bf Average} &            &    2:00:00 &    4:00:00 &    3:00:00 &                     \\

                                   \multicolumn{ 5}{l}{{\bf }} & {\bf High} &            &    4:00:00 &    8:00:00 &    6:00:00 &                    \\

    {\bf } &     {\bf } &     {\bf } &     {\bf } &     {\bf } &     {\bf } &            &            &            &            &                      \\

   \multicolumn{ 5}{l}{\cellcolor [gray]{0.9}{\bf Creation of CSS}} & {\bf Low} &            &    4:00:00 &    6:00:00 &    5:00:00 &                     \\

                                   \multicolumn{ 5}{l}{{\bf }} & {\bf Average} &            &   12:00:00 &   12:00:00 &   12:00:00 &                      \\

                                   \multicolumn{ 5}{l}{{\bf }} & {\bf High} &            &   24:00:00 &   24:00:00 &   24:00:00 &                     \\

    {\bf } &     {\bf } &     {\bf } &     {\bf } &     {\bf } &     {\bf } &            &            &            &            &                       \\

          \multicolumn{ 5}{l}{\cellcolor [gray]{0.9}{\bf Creation of JS/ASP functions}} & {\bf Low} &            &    3:00:00 &    3:00:00 &    3:00:00 &                      \\

                                   \multicolumn{ 5}{l}{{\bf }} & {\bf Average} &            &    5:00:00 &    5:00:00 &    5:00:00 &                       \\

                                   \multicolumn{ 5}{l}{{\bf }} & {\bf High} &            &    7:00:00 &    7:00:00 &    7:00:00 &                      \\

    {\bf } &     {\bf } &     {\bf } &     {\bf } &     {\bf } &     {\bf } &            &            &            &            &                     \\

 \multicolumn{ 5}{l}{\cellcolor [gray]{0.9} {\bf Adequation SQL, JS/ASP functions}} &  {\bf Low} &    &    2:30:00 &   2:30:00  &    2:30:00 &              \\
 \multicolumn{ 5}{l}{ {\bf }} &  {\bf Average} &     &  3:30:00 &   3:30:00 &   3:30:00 &               \\
\multicolumn{ 5}{l}{ {\bf }} & {\bf High} &      &   4:30:00 &   4:30:00 &   4:30:00 &      \\
 \\
    \multicolumn{ 5}{l}{\cellcolor [gray]{0.9} {\bf Creating SP/SQL and components}} & {\bf Low} &  &  4:30:00 &  4:30:00 & 4:30:00 &              \\

 \multicolumn{ 5}{l}{{\bf }} &  {\bf Average} &  &    6:30:00 &   6:30:00 &    6:30:00 &               \\

                                   \multicolumn{ 5}{l}{ {\bf }} & {\bf High} &  &   9:00:00 &  9:00:00 &   9:00:00 &             \\

    {\bf } &     {\bf } &     {\bf } &     {\bf } &     {\bf } &     {\bf } &            &            &            &            &                    \\

          \multicolumn{ 5}{l}{{\bf Adequation of procedure}} & {\bf Low} &            &    0:20:00 &    0:25:00 &    0:23:00 &                      \\

                                   \multicolumn{ 5}{l}{{\bf }} & {\bf Average} &            &    0:50:00 &    1:00:00 &    0:55:00 &                       \\

                                   \multicolumn{ 5}{l}{{\bf }} & {\bf High} &            &    1:50:00 &    2:20:00 &    2:00:00 &                      \\

    {\bf } &     {\bf } &     {\bf } &     {\bf } &     {\bf } &     {\bf } &            &            &            &            &                       \\

    \multicolumn{ 5}{l}{{\bf Creating maintenance page}} & {\bf Standard} &            &    0:05:00 &    0:10:00 &    0:08:00 &                     \\
    {\bf } &     {\bf } &     {\bf } &     {\bf } &     {\bf } &     {\bf } &            &            &            &            &                       \\

    \multicolumn{ 5}{l}{{\bf Survey with the client}} & {\bf Standard } &            &    24:00:00 &    36:00:00 &    32:00:00 &                    \\
    {\bf } &     {\bf } &     {\bf } &     {\bf } &     {\bf } &     {\bf } &            &            &            &            &                     \\

    \multicolumn{ 5}{l}{{\bf Site conversion to CMS}} & {\bf High} &            &    240:00:00 &    472:00:00 &    320:00:00 &                        \\

    {\bf } &     {\bf } &     {\bf } &     {\bf } &     {\bf } &     {\bf } &            &            &            &            &                     \\

\end{tabular}
\end{center}
\end{table*}

In sum, the classical FPA involves counting of type data functions and type transaction functions, such as inputs, outputs, inquiries, files and movements. Focusing only the functionalities associated with data-strong structures, the measurement based on function points at MTE does not recognize as worthy of metric criteria all the necessary interface architecture and presentation features, which ensure accessibility, friendly operability, as well as effective dissemination of subjects. 

\section{Methodology}
The basic premise of finding a criterion for counting the cases in question are the disadvantages of its own current, namely

\begin{itemize}
\item Worthlessness of professional specialization in the absence of a proper technique of counting and measuring, a fact which causes the depreciation of professional profile in the work market;
\item Exposure of the Coordination of IT to the questions of the organs of control because of the degree of arbitrariness in the calculation currently practiced;
\item Underestimation of intellectual effort, lowering the intangible value of hours worked;
\item Absorption of the fragility of the standard count done in other instances, insofar as most of these errors would be compensated by the low and always lowering cost of construction and development of presentation layers in the Extranet, Intranet or Internet;
\item Inaccurate billing from contractors.
\end{itemize}

Since the obvious disadvantages are shown for all, we deal with the necessary tools to seek the establishment of the technique described above. The first task undertaken was a survey, together with experts, comprising a consistent and clear roll of jobs related to the demands for the web area. The list, formatted in Table 1, was constructed keeping in mind the subsequent use of PERT - Program Evaluation and Review Technique - \citep{7}, thus containing the time intervals needed to perform jobs of low, medium and high complexity in optimistic, pessimistic and most likely perspectives. All convoked professionals, invited to construct the Table, are graphic designers (or industrial designers, with specialization in visual programming) with practice as query developers and have more than ten years experience in web development. They are professionals able to develop visual designs in various areas since the creation of corporate identity to several graphic pieces for both print and digital media. As we know, the area of web design requires the domain of important concepts of usability, navigation and web standards. Each professional (five in all) built his own Table, so that the final Table adopted computed the average of individual Tables. 

The complexity of the task is a linguistic variable assuming the fuzzy values "low", "average" and "high" \footnote{The fuzzy logic tries to fulfill the gap of infinite degrees of uncertainty between "to be" or "not to be". The intrinsic imperfections of the information represented in natural language have been treated more appropriately by fuzzy logic than by the theory of probabilities.}. In fact, fuzzy models are very useful in information technology such as those based on the creation of causal and cognitive maps of risks provided by experts experience \citep{6}. We note that in present model the complexity is classified by the duration of the task. The criteria for classification of complexity have been established by the respective groups of expert professionals selected to determine the times of PERT. For example, a high complexity query in SQL accesses more than three tables, while for a low complexity query only one table is accessed. Also the task list has not exhausted the range of all non-measurable items; insomuch, Table 1 may be expanded as necessary.  

Parallel to the construction of the Table, I have imagined that the theoretical number of function points for the cases not directly measurable would be, {\it a priori}, a function of time consumed and some feature of the development tool used, so that,
\begin{equation}
N = f(t,i),
\end{equation}
where $N$ is the number of function points, $t$ is the time taken for completion of the task and $i$ is called "inertia of development" or "constraint capacity", i. e., an index that represents the constraints imposed by the tool (Table 3). If desired, from the parameter $i$ we may define the productivity $p$ of the tool as,
\[
p = \frac{1}{i}.
\]
The values of $i$ were established with the aid of professional experience accumulated over the years of work and by reports on productivity from suppliers. Time is defined as in PERT, i. e.,
\begin{equation}
T = \frac{{O + 4MP + P}}{6},
\end{equation}
where $O$ is the optimistic estimate, $P$ is the pessimistic estimate and $MP$ is the more likely estimate. \\

\section{The systemic view of the software development}
As once told Mario Bunge (1961),\\ 

"One of the most difficult and interesting problems of rational decision is the choice among possible diverging paths in theory construction and among competing scientific theories, i.e., systems of accurate testable hypotheses. This task involves many beliefs-some warranted and others not as warranted and marks decisive crossroads.[...] The set of metascientific criteria dealing with the various traits of acceptable scientific theories is what guides the choice among competing courses in theory construction and among the products of this activity".\\\\
I usually say that it is easy to propose complicate things; difficult is to propose simple things. The set of metascientific criteria I adopted includes simplicity among the requirements that present model is supposed to satisfy. Here, the simplicity refers to semantical and pragmatical simplicities, that is, to an economy of additional meaning pre-suppositions and to an economy of work by who counts the system or project. I started from an econophysical vision, according to which a project is a system endowed with its own dynamics of evolution of human and material investments. As a physical system, and circumscribed by the development perspective, we have a Lagrangian function that describes this dynamic \citep{14}, taking into account the relevant parameters for the measurement of software projects. The simplest and typical Lagrangian form is given by,
\begin{equation}
\mathcal{L} = \frac {1}{2} N \dot N^2 - V,
\end{equation}
where the overdot indicates time derivative and $V$ is the potential of the project to generate function points. We note that Lagrangian $\mathcal{L}$ is not an explicit function of time. The function $\mathcal{L}$ is such that, 
\[
\frac{{d\mathcal{L}}}{{dt}} = \frac{{\partial \mathcal{L}}}{{\partial N}}\frac{{dN}}{{dt}} + \frac{{\partial \mathcal{L}}}{{\partial \dot N}}\frac{{d\dot N}}{{dt}}
\]
and obeys the Euler-Lagrange equation,
\begin{equation}
\frac{{d\mathcal{L}}}{{dt}} - \left[ {\frac{{dN}}{{dt}}\frac{d}{{dt}}\left( {\frac{{\partial \mathcal{L}}}{{\partial \dot N}}} \right) + \frac{{\partial \mathcal{L}}}{{\partial \dot N}}\frac{{d\dot N}}{{dt}}} \right] = 0,
\end{equation}
\[
\frac{{\partial \mathcal{L}}}{{\partial N}}\frac{{dN}}{{dt}} - \frac{{dN}}{{dt}}\frac{d}{{dt}}\left( {\frac{{\partial \mathcal{L}}}{{\partial \dot N}}} \right) = 0,
\]
\[
\frac{d}{{dt}}\left( {\frac{{\partial \mathcal{L}}}{{\partial \dot N}}} \right)\frac{{dN}}{{dt}} - \frac{{\partial \mathcal{L}}}{{\partial N}}\frac{{dN}}{{dt}} = 0,
\]
\[
\frac{d}{{dt}}\left( {\frac{{\partial \mathcal{L}}}{{\partial \dot N}}} \right) - \frac{{\partial \mathcal{L}}}{{\partial N}} = 0.
\]
Equation (4) may be rearranged and must determine a "conserved current" $J$ according to,
\[
\frac{d}{{dt}}\left[ {\mathcal{L} - \dot N\frac{{\partial \mathcal{L}}}{{\partial \dot N}}} \right] = 0,
\]
from which, 
\begin{equation}
\frac{{\partial \mathcal{L}}}{{\partial \dot N}}\dot N - \mathcal{L} = J = C^{te}.
\end{equation}
The quantity $dt$ is the time differential, $\dot N$ and $N$ are the generalized coordinates of the system. The derivatives of the Lagrangian are always taken with respect to the explicit generalized coordinates. As in physics the Lagrangian refers to energy units (such as $kg. m^2/s^2$), the analogical "conserved current" $J$ in present theory refers to the intellectual effort {\it per} squared hour. Applying equation (5) to function (3) we obtain,
\begin{equation}
\frac{1}{2}N\dot N^2  + V = J.
\end{equation}
Having conjectured and tested a few options of empirical formulas to match equation (6), I proposed the following more advanced expression as the best approach,
\begin{equation}
N = \frac{{C^i }}{{K(P-O+1)}}\left( {\frac{{O + 4MP + P}}{6}} \right),
\end{equation}
where $K$ is the contractual adjustment parameter, and $C$ is the intellectual effort conversion factor for function points according to Table 2. It is not the case of arbitrary statement but tacit assumption. A tacit assumption is performed from some rational premises of a logical argument and, by corroboration with experience, attains the status of a postulate. So, to be acceptable as postulate, tacit equation (7) has to be tested. Fixing $O$ and $P$ as time-extremes, we can do $MP=t$ (the time variable estimate). Adopting $K=1$ and inserting expression (7) into equation (6) we deduce,  
\[
\frac{1}{2}\frac{{C^i }}{{(P - O + 1)}}\frac{{(O + 4t + P)}}{6}\frac{{C^{2i} }}{{\left( {P - O + 1} \right)^2}}\times
\]
\begin{equation}
\frac{{16}}{{36}} + V = J.
\end{equation}
The "conserved current" $J$ is fixed around $4C^i/(P-O+1)^2$, so that,
\[
V = \frac{{4C^i }}{{(P-O+1)^2 }} - 
\]
\begin{equation}
\left[ {\frac{1}{2}\overbrace {\frac{{C^i }}{{(P - O + 1)}}\frac{{(O + 4MP + P)}}{6}}^N\underbrace {\frac{{C^{2i} }}{{\left( {P - O + 1} \right)^2 }}\frac{{16}}{{36}}}_{\dot N^2}} \right].
\end{equation}
Figure 1 shows the shape of theoretical curves of the potential $V$ according to the inertia of development $i$ for three values of $K$. A negative or even null potential may signify that the combination tool-effort-time is not being productive. Thereby, the potential can be an essential instrument to support project managers, providing a guide to better match the available resources.

We see that function (7) clearly satisfies the equation (4) and ensures current's units and time control of the number of function points, keeping the PERT time in the whole description (this is desirable, since the number of function points is very time-sensitive to the PERT estimate). The Lagrangian (3) represents the dynamics of the project along the time. The "conserved current" gives the effective intellectual effort {\it per} squared hour\footnote{The energy consumed by the intellectual effort during time $t$ has its equivalent in function points. The effort reduces, after all, to an amount of function points, so that the units of $\mathcal{L}$ and $J$ are the same.}.

To well understand my point in favor of exponential representation, it is necessary a brief look at the relation between exponential growth and geometric progression. If the ratio of a growth rate to the increasing quantity itself is a constant, we are dealing with a process that may be explained by an observational series of type $a, ak, ak^2, ak^3,..., ak^n$, says, a geometric progression of ratio $k$. Such increasing quantity, $Q$, being considered as a continuous function of time, $t$, may be dynamically described by means of a differential equation of type $dQ/dt=rQ$ for $a = Q(0)$, to what the unique solution is $Q = a.exp(rt)$. Thus, any set of measurements of the quantity $Q$ for $t = 0, 1, 2, 3,…$ is a geometric progression where $k = e^r$. As we see, exponential growth refers to a continuous natural change in which multiplication is fractally-repeated; geometric progression is a discrete subset of that continuous \citep{24}. In science, the use of exponentiation is pervasive in several fields, mainly in economics and physics.
\begin{figure} [h]
\includegraphics[scale=0.336]{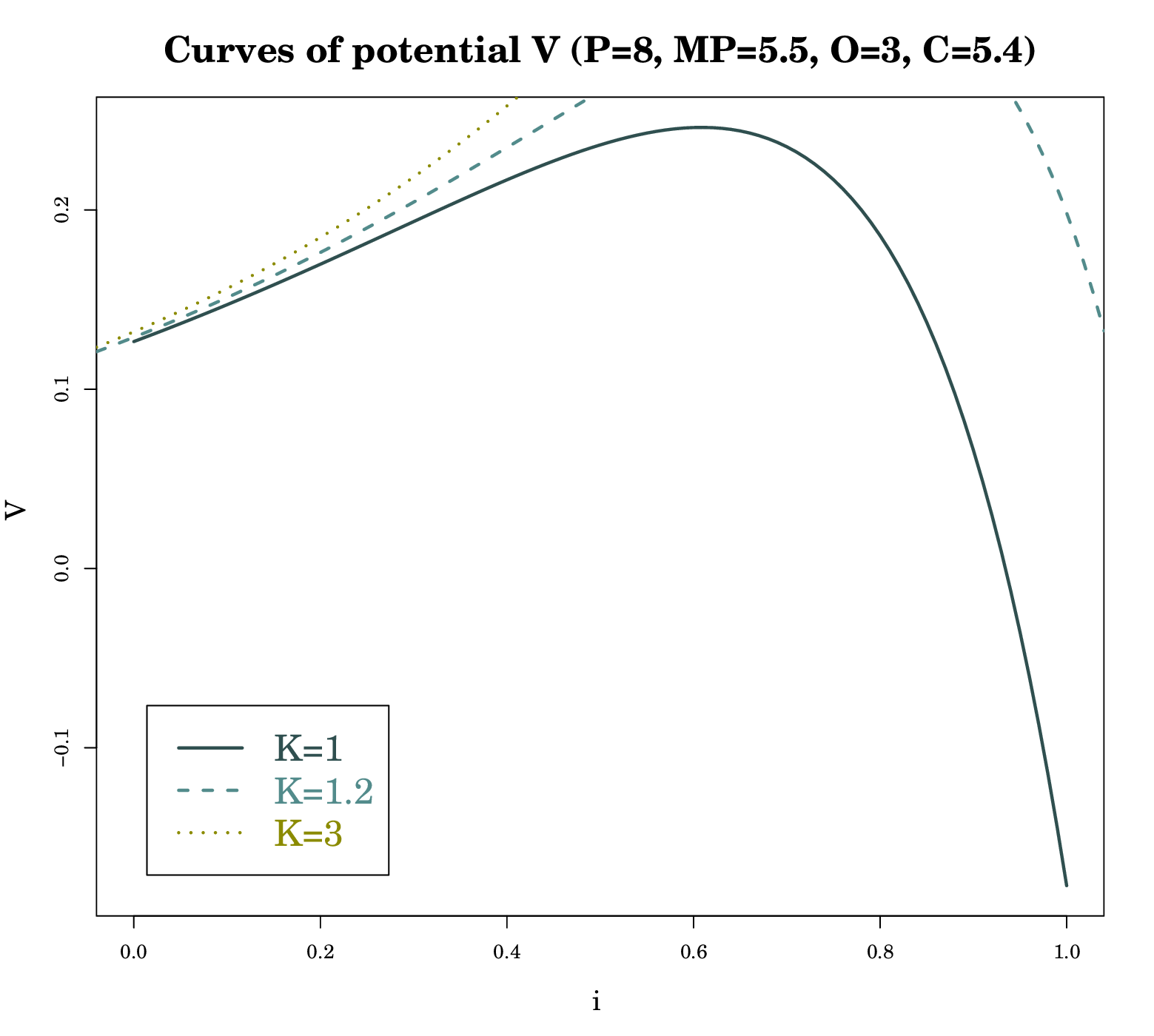}
\\
\small {Figure 1: {\it Theoretical curves of the potential $V$ ranging from $i=0$ to $i=1$ for different values of $K$.}} 
\end{figure}

Now, the counting of function points must be a continuous process, since precision is a pivotal claim to pay for a task. As exponentiation grows faster than multiplication, the first is useful to describe a quantity that must evolve more quickly beneath the weight of another quantity that represents the ever increasing performance of the modern tools. Yet, the intellectual effort is obviously very influenced by the tool in use, so that it is reasonable to accept the factor $C^i$.

Since function points are now being applied far beyond the basic purpose of a sizing mechanism for software projects, including litigation involving software contracts and software taxation, the parameter $K$ enters to prevent problems already in the early stages of negotiation, adjusting the range of the counting to the provisos of a given contract. 

There are numerous conversion Tables in the market. Tables (2) and (3) purposely present fractal weights to converting time in function points, since at the end we want to find monetary values. The application of weights is very known from fuzzy logic \citep{3} to quantify intuitive concepts as "intellectual effort" and "inertia of development"; here they were defined also in comparison with similar Tables used for the reverse path, i. e., for the transformation of function points in hours worked \citep{10}, taking into account the level of intellectual investment in each phase. Just as experts opine on the hierarchy of the project phases, they also opine on the degree of intellectual relevance, from 1 to 6, adopted for each component phase of the project, as presented in Table 2, including the elements with higher weights at higher effort. The range of the weights was delimited to confine the number of function points within acceptable intervals. Also, informations provided by suppliers were used in some cases (Table 3). For instance, a certain supplier of CMS tools reported that Lumis\footnote{The Lumis is a Brazilian software company pioneering in development of products and solutions for enterprise portals.} gives a gain of thirty percent in productivity compared to older similar and less friendly applications. This means that $i$ is equal to $1-0.3=0.7$, being $1$ the weight of the hardest tool adopted as reference to compare the power of Lumis. In other words, Lumis adjusts intellectual effort from $C$ to $C^{0.7}$. The contractual adjustment parameter $K$ regulates the growth of the number of function points according to the budget levels for each institution, keeping scores on appropriate scales.  

\begin{table}[t]
\caption{Converting factors for function points}
\begin{tabular}{@{}lr}
\hline

{\bf Step} & {\bf Factor} \\

Survey &       3.2  \\

Elaboration &       5.8  \\

Construction &       (**) \\

    Tests &       2.6 \\

    Alteration &       1.5 \\

Implantation &        1.2  \\

           &            \\

           &            \\

(**) Languages &            \\

{\bf Internet/Intranet} & {\bf Factor} \\

PHP, Java Script and ASP &    3.5 \\

      HTML &          1.8  \\

Java, CMS tools, ETL &       5.4  \\

\hline
\end{tabular}
\end{table}

\begin{table}[t]
\caption{Inertia of development $i$ for some tools}
\begin{tabular}{@{}lr}
\hline

{\bf Tool} & {\bf $i$} \\

HTML &                   1.00  \\

ASP  &                  0.90  \\

CMS tools (Lumis, Vignette, etc.), PHP &  0.70  \\

Statistics tools &     0.65 \\

Management tools &     0.60 \\

DBA tools (Query Analyzer, V. Studio, etc.) &   0.52 \\

Text processors   &    0.60 \\

Text editors     &     0.50 \\

Development tools (Photoshop, Corel, etc.) &            0.68 \\

Humanware        &    0.00 \\

ETL   &            0.36 \\

           &            \\
\hline
\end{tabular}
\end{table}

\begin{figure} [h]
\includegraphics[scale=0.336]{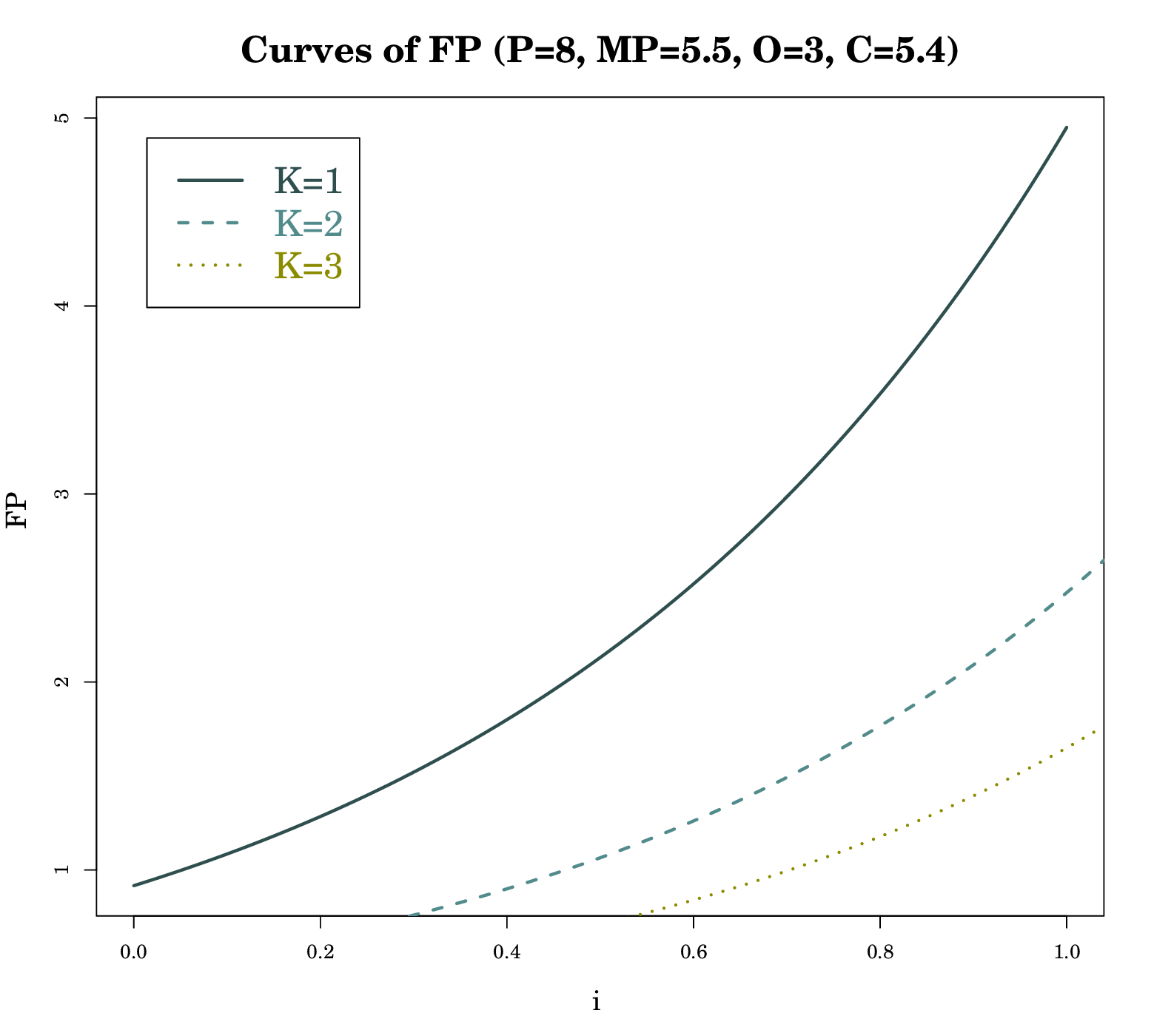}
\\
\small {Figure 2: {\it Theoretical curves of function point counts ranging from $i=0$ to $i=1$ for different values of $K$. Note that for $K >1$ the amount of function points decreases.}} 
\end{figure}
\begin{figure} [h]
\includegraphics[scale=0.336]{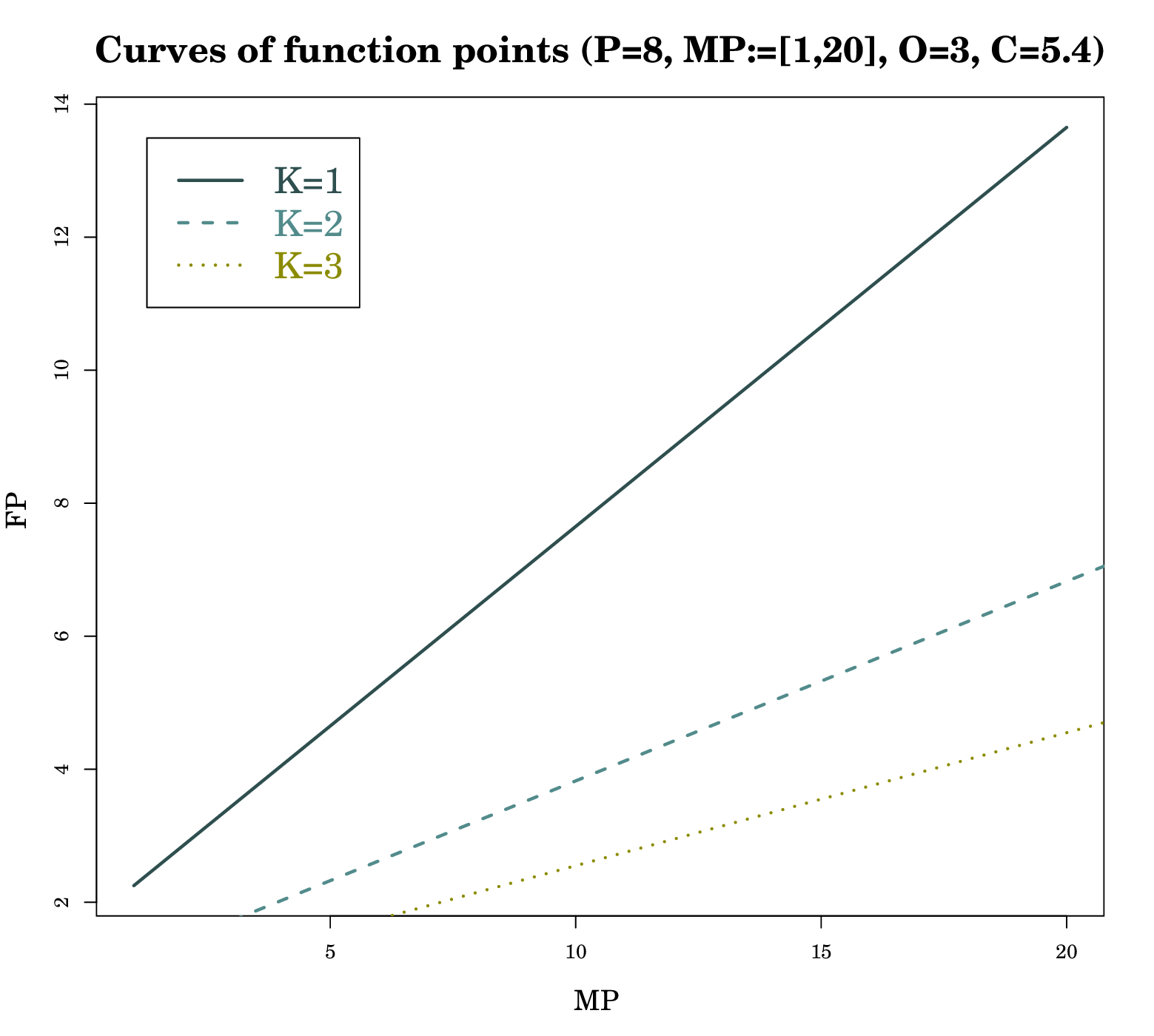}
\\
\small {Figure 3: {\it Theoretical curves of function point counts ranging from  $MP=1$ to $MP=20$ and $i=0$ to $i=1$ for different values of $K$.}} 
\end{figure}

\section{Results}
Let us understand in details the formula (7). Mathematics changes to number, by own syntax, a clause written in plain language. Unfortunately, the teaching of the subject does not always makes this clear, so that the mathematical expressions associated with empirical processes often seem obscure. So, by the requirement of semantical simplicity, the simple clause, \\\\
"The number of function points is directly proportional to the intellectual effort ($C$) powered by the constraint capacity of the tool ($i$) and to the estimated average time to complete the task, and inversely proportional to the difference between the optimistic and the pessimistic ending outlooks",
\\\\
means in symbolic language nothing more than equation (7), says,  
\[
N = \frac{{C^i }}{{K(P - O + 1)}}\left( {\frac{{O + 4MP + P}}{6}}\right).
\]
In other words, the total score increases with the time consumed and the intellectual effort, but decreases as the tool is more productive. Of course, the more the tool optimizes the work, the more the cost in intellectual effort is pared ($C^i$), and we may look at parameter $K$ as a constant of proportionality. The time variance is given by,
\begin{equation}
\sigma ^2  = \left( {\frac{{P - O}}{6}} \right)^2, 
\end{equation}
where $\sigma$ is the standard deviation. In terms of the deviation $\sigma$ we may write,
\begin{equation}
N = \frac{{C^i }}{{K(6\sigma  + 1)}}\left( {\frac{{O + 4MP + P}}{6}} \right).
\end{equation}
Note that when two or more quantities are independent degrees of freedom that interact with one another, we multiply them (in present case, time, inertia and effort). Given that the PERT factor has temporal dimension and tends to increase as the complexity of the work, to slow the correlative increase of the score points and to obtain a dimensionless number that can be interpreted as the number of function points it is needed a denominator with relevant time content. The difference between pessimistic and optimistic estimates plus one, $P-O+1$, provides the maximum time span within which the work will be consummated; it enters the formula with weight $K = 1$ based on the given IT current contract at MTE to better fit the economic reality of this Ministry, with the addition of 1 ensuring that the denominator shall never be zero if the two estimates are identical. This means that the total score decreases as the optimistic time gets away from the pessimistic outlook, a manner to provide a kind of "discount" in function points from the delay of task completion (uncertainty in the ending time of the task). Indeed, the idea of a discount in function points by the "uncertainty" in the delivery time was accepted as a fair criterion in face of the ever-present urgency of the user. It is convenient to recall here that the conversion factor $C$ accounts for the intellectual weight of each task and the inertia $i$ tells to us whether the effort will be greater or smaller according to the chosen tool. When no software tool is used, i. e., the activity is performed only with humanware, the value of the inertia of development is zero. This applies to raising activities in which employees are not in use of softwares for management and monitoring of projects, spreadsheets and others, being the work organized in notes and handmade schedules. This exponent covers the range $[0,1]$, being highest in the HTML markup language, taken as a  more constraining technological alternative for web. The inertia was fixed within the class of each tool. Figure 2 shows the shape of theoretical curves of counting according to the inertia of development $i$ for three values of $K$. Figure 3 shows the shape of theoretical curves of counting in a different perspective, according to the simultaneous variation of time $MP$ and  inertia of development $i$ for the same three values of $K$. 

An interesting question appears when no tool is used, such that, for $i = 0$, there is always $C$ powered by $0$, i. e., $C^0 = 1$. Thus, the intellectual effort would be, by definition, equal to $1$ in any situation in which no tool was required. In this case, the conclusion is simply that without the use of any tool it would be largely arbitrary to differentiate intellectual efforts, since the idea is precisely to evaluate changes in effort in the presence of a tool in a certain project phase. Therefore, if no tool is applied, we normalize the effort for all individuals and project phases, so that it does not depend on particular task or individual itself. In this case, the difference in function points is in charge of the PERT time and variance $P - O$ (indeed, it is logically consistent with the fact that without tool there is no "measurement apparatus" that allows to evaluating the intellectual effort by the interaction man-tool). For instance, let us take the hypothetic counting of a survey with the client not supported by any software. From Table 1 we have the following values:

\begin{itemize}
\item $O=24:00h$,
\item $P=36:00h$,
\item $MP=32:00h$.
\end{itemize}
From Table 2 we extract for $C$ the value $3.2$ corresponding to the survey step. Since there is no tool, $i=0$. Then, formula (7) computes,
\[
N = \frac{{1}}{{13}}\left( {\frac{{24 + 4 \times 32 + 36}}{6}} \right) \simeq 2.41.
\]
Taking the value of the function point for new implementations, $V_p=$R\$$480.00$, the total value to pay by the task would be $V_t=$R\$$1,156.92$. Now, supposing the use of management tools ($i=0.60$), we get, 
\[
N = \frac{{3.2^{0.60}}}{{13}}\left( {\frac{{24 + 4 \times 32 + 36}}{6}} \right) \simeq 4.84,
\]
amount of points that multiplied by the point value furnishes $V_t=$R\$$2,324.84$. The difference in values means that between using and not using a tool there is a logical difference of intellectual effort; as well as the knowledge required for the survey, it is necessary to know the specific software.

Now we will see some examples of calculation, from the experience in MTE, to validate equation (7) as a postulate. Be the task of drafting a new form of medium complexity. From Table 1 we have the following values:

\begin{itemize}
\item $O=2:00h$,
\item $P=8:00h$,
\item $MP=6:00h$.
\end{itemize}
From Table 2 we extract for $C$ the value $5.8$ corresponding to the elaboration step. Since the tool is the HTML markup language, $i=1$. Then, formula (7) computes,
\begin{equation}
N = \frac{{5.8}}{{7}}\left( {\frac{{2 + 4 \times 6 + 8}}{6}} \right) = 4.6952.
\end{equation}
Taking the value of the function point for new implementations, $V_p=$R\$$480.00$, the total value to pay by the task would be $V_t=$R\$$2,253.70$. For a form of high complexity, have done at the same conditions,
\begin{itemize}
\item $O=4:00h$,
\item $P=12:00h$,
\item $MP=9:00h$,
\end{itemize}
\begin{equation}
N = \frac{{5.8}}{{9}}\left( {\frac{{4 + 4 \times 9 + 12}}{6}} \right) = 5.5852,
\end{equation}
amount of points that multiplied by the point value furnishes $V_t=$R\$$2,680.89$. Note that the difference in amount to be paid is controlled by the denominator with weight 1 times the expanded time-range $P-O+1$.

Finally, the example of the creation/development of a layout. For high complexity we get from Table 1,
\begin{itemize}
\item $O=38:00h$,
\item $P=40:00h$,
\item $MP=39:00h$,
\end{itemize}
values which introduced in formula (7) during the construction phase in HTML provide,
\begin{equation}
N = \frac{{1.8}}{{3}}\left( {\frac{{38 + 4 \times 39 + 40}}{6}} \right) = 23.40,
\end{equation}
amount of points that multiplied by the point value gives $V_t=$R\$$11,232.00$. 

A caution to be taken, however, refers to whether the form is, even being new, a new implementation or can be considered within the context of an evolving maintenance in the body of a broader set of objects belonging to the same application. If it is new, but belonging to a previously developed application in another area, it will be considered new implementation; if it is new, but belonging to a previously developed application in the same area, it will be considered evolving maintenance.

\subsection{{\bf The counting of the Projective Statistical Bulletin}}

The Projective Statistical Bulletin is a real quarterly publication of the MTE, made available on the Intranet in "pdf" format with online abstract. Thus, it is not an web application, so it is necessary to quantify the parametric variable $i$ in terms of inertia of development of statistical modelling tools in use. Technically, the editors of the bulletin assume that $ i = 0.65$, being the intellectual effort punctuated by Table 2, item "Elaboration", i. e., $ C = 5.8$. Regarded as a highly complex product, it provides the following steps:
\begin{enumerate}
\item Survey of variables
\begin{itemize}
\item $O=20:00h$,
\item $P=30:00h$,
\item $MP=25:00h$,
\item $i = 0,7$ (search in statistical databases),
\item $C = 3,2$ (survey).
\end{itemize}
Thus, expression (7) computes
\begin{equation}
N = \frac{{3,2^{0,7} }}{{11}} \times \left( {\frac{{20 + 100 + 30}}{6}} \right)=
\end{equation}
\[
=0,205 \times 25 = 5,13.  
\]
\item ETL process
\begin{itemize}
\item $O=50:00h$,
\item $P=80:00h$,
\item $MP=65:00h$,
\item $i = 0,36$ (ETL tool),
\item $C = 5,4$ (ETL language).
\end{itemize}
Thus, expression (7) computes
\begin{equation}
N = \frac{{5,4^{0,36} }}{{31}} \left( {\frac{{50 + 260 + 80}}{6}} \right) = 
\end{equation}
\[
=0,059 \times 65 = 3,85.
\]
\item Statistical analysis
\begin{itemize}
\item $O=50:00h$,
\item $P=80:00h$,
\item $MP=65:00h$,
\item $i = 0,65$ (statistical tools),
\item $C = 5,8$ (elaboration).
\end{itemize}
Thus, expression (7) computes
\begin{equation}
N = \frac{{5,8^{0,65} }}{{31}} \times \left( {\frac{{50 + 260 + 80}}{6}} \right) = 
\end{equation}
\[
=0,101 \times 65 = 6,57.
\]
\item Text elaboration 
\begin{itemize}
\item $O=40:00h$,
\item $P=60:00h$,
\item $MP=50:00h$,
\item $i = 0,40$ (Word tool),
\item $C = 5,8$ (elaboration).
\end{itemize}
Thus, expression (7) computes
\begin{equation}
N = \frac{{5,8^{0,4} }}{{21}} \times \left( {\frac{{40 + 200 + 60}}{6}} \right) = 
\end{equation}
\[
=0,096 \times 50 = 4,81.
\]
\end{enumerate}

The total function points is, 
\[
N_{Total}  = 5,13 + 3,9 + 6,57 + 4,81 = 20,41,
\]
amount that multiplied by the point value for development ($V_p=$R\$$480.00$) produces $V_t=$R\$$9,796.80$. This example is sufficient to give an idea of how to apply the Tables in scoring a variety of tasks.

\subsection{Improper counting}
In some cases, when the estimate equals the optimistic and pessimistic, we have the "aberration count", that is, an amount of function points higher than the initial average score expected. In these situations, I recommend to assign a maximum similarity of 80\% between the two estimates; in other words, the pessimistic estimate should exceed the  optimistic at least 20\% of the latter to maintain the objectivity of the equation (7). Let us take the example of creation/development of low complexity layout. By Table 1,

\begin{itemize}
\item $O=8:00h$,
\item $P=8:00h$,
\item $MP=8:00h$.
\end{itemize}
These values entered in the formula (7) during the construction phase in HTML provide,
\begin{equation}
N = \frac{{1.8}}{{2,6}}\left( {\frac{{8 + 4 \times 8.8 + 9.6}}{6}} \right) = 6.092,
\end{equation}
amount of points that multiplied by the point value gives $V_t=$R\$$2,924.30$. If this were not done, 
\begin{equation}
N = \frac{{1.8}}{{1}}\left( {\frac{{8 + 4 \times 8 + 8}}{6}} \right) = 14.4,
\end{equation} 
amount of points that multiplied by the point value furnishes $V_t=$R\$$6.912,00$, a result that, by good sense, would be much more suitable for a layout of medium complexity. Here, we see how the number of function points is strongly sensitive to the duration of the task. Therefore, considerable degree of caution is part of the counting of function points, even in the classical approach of the technique. The reader must note that the prescribed minimum difference percentage of 20 \% between optimistic and pessimistic perspectives of task completion was established to address the particular situations (not to all non-measurable tasks) in which professionals can not discern objective estimates of duration, whether by level of control over the work, whether by features of the tool, or even by total dedication to the demand (it is common among programmers to devote exclusively to a certain problem until it is solved). Thus, the corresponding 20 \% variance, enough to ensure significant results in the present metric, is primarily applicable to strong-function items such as the reader can infer from a quick look at Table 1 (this Table highlights in grey background the typical tasks where occurs aberration counts). Table 4 shows a real counting at MTE with columns having point values for the tasks and deflation factors according to the current contract (the function point for adequation is more expensive). This counting applied the 20 \% variance between optimistic and pessimistic estimates.

There is also the issue of high number of pages, images and others in the development of sites and portals, which can distort considerably the final value to be paid by extrapolating at very fair the payment for the required work. Unless the so-called "maintenance pages", for the management of content for all other development items (images, pages, videos, etc.) I suggest the following rule:

\begin{flushleft}
\begin{enumerate}
\item number of pages, images, etc. between $1$ \\and $9$ $ \; \Rightarrow \; Quantity = 1$;
\item number of pages, images, etc. $\geq 10$ and $< 60 \Rightarrow \; Quantity = 2$;
\item number of pages, images, etc. $\geq 60$ and $< 100 \; \Rightarrow \; Quantity = 3$;	
\item number of pages, images, etc. $\geq 100$ and $< 500  \; \Rightarrow \; Quantity = 4$;
\item number of pages, images, etc. $\geq 500$ and $< 1000  \; \Rightarrow \; Quantity = 8$;
\item number of pages, images, etc. $\geq 1000 \; \Rightarrow \; Quantity = 16$.
\end{enumerate}
\end{flushleft}

\subsection{The counting as continuous function of the time}
Strictly speaking, a project constitutes a typical nonlinear system. However, for all practical purposes, it is reasonable to do a simple linear approach for the time evolution of the number counts, applying {\it a posteriori} corrections in order to minimize the effects of unpredictable fluctuations.

The theoretical count of function points is a continuous function of the most likely time $MP$. This is a simple conclusion, if we think that both $O$ and $P$ are fixed, being $MP$ widely variable. While $i$ and $C$ are usually regarded as parameters, $i$ can be taken by a flux between $0$ and $1$ (figures 2 and 3) if we think that the productivity of a tool varies continuously with practice stemming from the frequent use. Although a good way to visualize the shape functions, it is a difficult and poorly pragmatic approach (to quantify the increased productivity of the tool according to usage is a process that ends, for all purposes, in the reduction of observational variable $MP$).

In the case of time, the derivative of $N$ with respect to $MP$ will give us the increase or decrease in the number of function points {\it per} hour difference in $MP$. Thus, the equation
\begin{equation}
\frac{{dN}}{{dMP}} = \frac{{2C^i }}{{3K\left( {P - O + 1} \right)}}
\end{equation}
provides the number of points per hour to be added or subtracted from the total initial target after completion of the task, when we register the performed range of $MP$. Thus, the final number of function points is given by the gauge function,
\begin{eqnarray}
N \pm \frac{{dN}}{{dMP}} = \frac{{C^i }}{{K\left( {P - O + 1} \right)}}\left( {\frac{{O + 4MP + P}}{6}} \right) \pm 
\end{eqnarray} 
\[
 \pm \frac{{2C^i H}}{{3K\left( {P - O + 1} \right)}},
\]
\begin{equation}
\mathord{\buildrel{\lower3pt\hbox{$\scriptscriptstyle\smile$}} 
\over N}  = \frac{{C^i }}{{K\left( {P - O + 1} \right)}}\left( {\frac{{O + 4MP + P \pm 4H}}{6}} \right),
\end{equation}
where ${\mathord{\buildrel{\lower3pt\hbox{$\scriptscriptstyle\smile$}} 
\over N} }$ is the gauged number of function points and $H$ is the number of hours to sum or subtract from $MP$. Equation (23) is very useful to adjusting function points for simple jobs with no need of new additional counts. In these cases, it is enough to calibrate the number of hours ($MP$) to fit the required payment.

\begin{sidewaystable}\footnotesize
\caption{Tasks, complexities and durations for one real counting at MTE}
\begin{tabular}{lrrrrrrrrrrrrr}
\hline
\hline
{\bf Counting} & {\it C} & {\bf Compl} & {\bf O(h)} & {\bf MP(h)} & {\bf P(h)} & {\it i} & {\bf Points/unity} & {\bf Point value} & {\bf Amount} & {\bf Defl} & {\bf Tot value} & {\bf Defl value} & {\bf Tot points} \\
\hline
\hline
{\bf 1- System CPMR} &            &            &            &            &            &            &            &            &            &            &   {\bf   } &    {\bf  } &            \\

1.1- Procedure adequation &        3.5 &       high &       1.83 &       1.88 &       2.33 &       0.52 &     2.4895 &     537.00 &        100 &          2 & {\bf 133,687.25} & {\bf 66,843.63} & {\bf 248.95} \\

1.2- SQL adequation &        3.5 &       high &        3.6 &          4 &        4.5 &       0.52 &     4.0553 &     537.00 &        150 &          2 & {\bf 326,657.86} & {\bf 163,328.93} & {\bf 608.30} \\

1.3- JS creation &        3.5 &      average &          4 &        4.2 &          5 &       0.52 &     4.1243 &     480.00 &         30 &          1 & {\bf 59,390.41} & {\bf 59,390.41} & {\bf 123.73} \\

1.4- Component creation &        3.5 &       high &          4 &        4.2 &          5 &       0.52 &     4.1243 &     480.00 &         30 &          1 & {\bf 59,390.41} & {\bf 59,390.41} & {\bf 123.73} \\

1.5- ASP adequation &        3.5 &       high &        3.6 &        3.8 &        4.5 &       0.52 &     3.9207 &     537.00 &        180 &          2 & {\bf 378,977.34} & {\bf 189,488.67} & {\bf 705.73} \\

1.6- ASP creation &        3.5 &       high &        5.6 &        5.8 &          7 &       0.52 &     4.7691 &     480.00 &        160 &          1 & {\bf 366,266.45} & {\bf 366,266.45} & {\bf 763.06} \\

{\bf TOTAL} &            &            &            &            &            &            & {\bf } &            &            &            & {\bf 1,324,369.73} & {\bf 904,708.50} & {\bf 2,573.50} \\
\hline
\hline

\end{tabular}
\end{sidewaystable}

\section{Conclusion and final remarks}
This study presented an extended technique of measurement by function points for the so-called non-measurable items. It showed how to apply this technique to different examples of the everyday tasks in a real IT area, the General Coordination of Information and Informatics of the Ministry of Work and Employment at Brazil. The use of this technique, now included in our Systems Development Methodology, has proven to be compatible with local market practices and budgeting constraints normally present in the public administration.

Clearly, no counting technique is perfect. With practice, we see that some counts will produce higher values, some lower values, in a dialectic that sometimes seems unfair, but that at the final will show equilibrium of the overall balance billed by a simple tradeoff between the individual amounts. As Anita Cassidy
and Keith Guggenberger say, "Metrics should be viewed as navigational data rather than as conclusions or destinations" \citep{9}. It should be noted that the Tables may increase at any time due to possible needs for adaptation to unforeseen situations. I am firmly convinced that the technique works, both by the relative success it has achieved as the logic upon which it was built.

Anyway, quality is an issue that permeates all activities and their outcomes, including methodologies and techniques of calculation. As can be seen, the introduction of the theoretical expression (7) for calculating function points relating to the so-called non-measurable items does not want to maximize or minimize the amounts involved. This equation merely establish a fair and logical criterion of valuation, based on knowledge embodied in the market, in such a way as to provide managers with tools technically well made and safe, and to include in metric contractual features, with the same importance and seriousness, so essential IT tasks such as creation and development of the shape of our sites and portals.

\section{Acknowledgements}
The author acknowledges the IT Manager of the Ministry of Work and Employment at Brazil, Mr. Sergio Alves Cotia, for the comments on the manuscript and for his integral support to this work. The author is also grateful for the encouragement received from fellows certified by IFPUG. 
$$
$$
\begin{center}
$\diamondsuit\diamondsuit\diamondsuit$
\end{center}

\vspace{6.0in}

\def\notesname{Note}

\theendnotes


\newpage

\begin{thebibliography}{4}
\bibitem[\protect\citeauthoryear{Abrah\~{a}o \& Pastor}{2003}]{1}
Abrah\~{a}o, S. $\&$ Pastor, O. (2003). Measuring the functional size of web applications. International Journal of Web Engineering and Technology,  vol1,  1, 5-16. 
\bibitem[\protect\citeauthoryear{Abran \& Robillard}{1994}]{2}
Abran, A. $\&$ Robillard, N. (1994). Function points: A study of their measurement processes and scale transformations. J. Systems Software, 25, 171-184.
\bibitem[\protect\citeauthoryear{Aguiar}{1999}]{3}
Aguiar, H. (1999). L\'{o}gica difusa: Aspectos pr\'{a}ticos e aplica\c{c}\~{o}es. Editora Interci\^{e}ncia, Brasil.
\bibitem[\protect\citeauthoryear{Albrecht \& Gaffney}{1983}]{4}
Albrecht, A. $\&$ Gaffney, J. (1983). Software function, source lines of code, and development effort prediction: A software science validation. IEEE Trans. Softw. Eng., 9(6):639–648.
\bibitem[\protect\citeauthoryear{Aramo}{2011}]{5}
Aramo-Immonen, H. {\it et al} (2011). Multicultural software development: The productivity perspective. International Journal of Information Technology Project Management, 2(1), 19-36.
\bibitem[\protect\citeauthoryear{Bodea \& Dascalu}{2010}]{6}
Bodea, C. N. \& Dascalu, M. I. (2010). IT risk evaluation model using risk maps and fuzzy inference. International Journal of Information Technology Project Management, v1, 79-97.
\bibitem[\protect\citeauthoryear{Boiteux}{1985}]{7}
Boiteux, C. D. (1985). PERT/CPM/ROY e outras t\'{e}cnicas de programa\c{c}\~{a}o e controle. Livros T\'{e}cnicos e Cient\'{i}ficos. Brasil.
\bibitem[\protect\citeauthoryear{Bunge}{1961}]{8}
Bunge, M. (1961). The weight of simplicity in the construction and assaying of scientific theories.  Philosophy of Science, 28, 260-281.
\bibitem[\protect\citeauthoryear{Cassidy \& Guggenberger}{2001}]{9}
Cassidy, A. \& Guggenberger, K. (2001). A practical guide to information systems process improvement.  St. Lucie Press - Boca Raton $\bullet$ London $\bullet$ Washington, D. C. $\bullet$ New York. 
\bibitem[\protect\citeauthoryear{CTIS}{2004}]{10}
CTIS (2004). Manual de contagem de ponto de fun\c{c}\~{a}o. Departamento de Projetos, Divis\~{a}o de F\'{a}brica de Software. Bras\'{i}lia, Brasil.
\bibitem[\protect\citeauthoryear{Drach}{2005}]{11}
Drach, M. (2005). Aplicabilidade de m\'{e}tricas por pontos de fun\c{c}\~{a}o a sistemas baseados em Web. Professional Master Thesis. Instituto de Computa\c{c}\~{a}o. Unicamp, Brasil.
\bibitem[\protect\citeauthoryear{Dreger}{1989}]{12}
Dreger, J. (1989). Function point analysis. Prentice Hall, Englewood Cliffs, New Jersey.
\bibitem[\protect\citeauthoryear{Garmus \& Herron}{2001}]{13}
Garmus, D. \& Herron, D. (2001). Function point analysis – measurement practices for successful software projects. Addison-Wesley, Massachusetts.
\bibitem[\protect\citeauthoryear{Goldstein {\it et al}}{2001}]{14}
Goldstein, H.  {\it et al} (2001). Classical mechanics.  Addison Wesley, Massachusetts.
\bibitem[\protect\citeauthoryear{Hern\'{a}ndez-L\'{o}pez {\it et al}}{2011}]{15}
Hernández-López, A. {\it et al} (2011). Software engineering productivity: Concepts, issues and challenges. International Journal of Information Technology Project Management, 2(1), 37-47.
\bibitem[\protect\citeauthoryear{IFPUG}{2000}]{16}
IFPUG (2000). Medi\c{c}\~{a}o de pontos por fun\c{c}\~{a}o a partir da especifica\c{c}\~{a}o orientada a objeto.  Function Point Counting Practices Manual, Versão 4.1.1. Retrieved January 16, 2010, from  http://www.ifpug.org.
\bibitem[\protect\citeauthoryear{Jalote}{1998}]{17}
Jalote, P. (1998). An integrated approach to software engineering. Narosa Publishing House, New Delhi. 
\bibitem[\protect\citeauthoryear{Jones}{1994}]{18}
Jones, C. (1994).  Software metrics: Good, bad and missing. Computer, 27(9): 98–100.
\bibitem[\protect\citeauthoryear{Kemerer}{1993}]{19}
Kemerer, C. (1993). Reliability of function points measurement – a field experiment. Communications of the ACM, 36(2): 85–97.
\bibitem[\protect\citeauthoryear{Lokan}{2008}]{20}
Lokan, C. J. (2008). Function points.  School of Information Technology and Electrical Engineering, Camberra.
\bibitem[\protect\citeauthoryear{Micro Focus}{2008}]{21}
Micro Focus Enterprise View (2008). Technical function point. Micro Focus. Retrieved January 22, 2010, from http://supportline.microfocus.com.
\bibitem[\protect\citeauthoryear{Mustafa {\it et al}}{2005}]{22}
Mustafa, K. {\it et al} (2005). Measuring the function points for migration project: A case study. American Journal of Applied Science, 2 (8), 1218-1221.
\bibitem[\protect\citeauthoryear{Ram {\it et al}}{2000}]{23}
Ram, D. {\it et al} (2000). An approach for pattern oriented software development based on a design handbook. Annals of Software Engineering, 10.
\bibitem[\protect\citeauthoryear{Serpa}{2005}]{24}
Serpa, N. (2005). Simple approaches for demographic estimations during health projective studies. In:  IX World Congress on Health Information and Libraries – VII Latin American and Caribean Congress on Health Sciences Information, ICML9 Abstracts, Brasil.
\bibitem[\protect\citeauthoryear{Whitmire}{1993}]{25}
Whitmire, S. (1993). Applying function points to object oriented software. Software Engineering Productivity Handbook, McGraw-Hill.
\end{thebibliography}
\end{document}